
\magnification=\magstep1
\font\bm=cmbx12
\font\tiny=cmr8
\def\rr{{\bf r}}


%
\catcode`@=11
%
%
\def\b@lank{ }

\newif\if@simboli
\newif\if@riferimenti

\newwrite\file@simboli
\def\simboli{
    \immediate\write16{ !!! Genera il file \jobname.SMB }
    \@simbolitrue\immediate\openout\file@simboli=\jobname.smb}

\newwrite\file@ausiliario
\def\riferimentifuturi{
    \immediate\write16{ !!! Genera il file \jobname.AUX }
    \@riferimentitrue\openin1 \jobname.aux
    \ifeof1\relax\else\closein1\relax\input\jobname.aux\fi
    \immediate\openout\file@ausiliario=\jobname.aux}

\newcount\eqnum\global\eqnum=0
\newcount\sect@num\global\sect@num=0

\newif\if@ndoppia
\def\numerazionedoppia{\@ndoppiatrue\gdef\la@sezionecorrente{\the\nchap}}

\def\se@indefinito#1{\expandafter\ifx\csname#1\endcsname\relax}
\def\spo@glia#1>{} 

\newif\if@primasezione
\@primasezionetrue

\def\s@ection#1\par{\immediate
    \write16{#1}\if@primasezione\global\@primasezionefalse\else\goodbreak
    \vskip\spaziosoprasez\fi\noindent
    {\bf#1}\nobreak\vskip\spaziosottosez\nobreak\noindent}
%

\def\sezpreset#1{\global\sect@num=#1
    \immediate\write16{ !!! sez-preset = #1 }   }

\def\spaziosoprasez{50pt plus 60pt}
\def\spaziosottosez{15pt}

\def\sref#1{\se@indefinito{@s@#1}\immediate\write16{ ??? \string\sref{#1}
    non definita !!!}
    \expandafter\xdef\csname @s@#1\endcsname{??}\fi\csname @s@#1\endcsname}

\def\autosez#1#2\par{
    \global\advance\sect@num by 1\if@ndoppia\global\eqnum=0\fi
    \xdef\la@sezionecorrente{\the\sect@num}
    \def\usa@getta{1}\se@indefinito{@s@#1}\def\usa@getta{2}\fi
    \expandafter\ifx\csname @s@#1\endcsname\la@sezionecorrente\def
    \usa@getta{2}\fi
    \ifodd\usa@getta\immediate\write16
      { ??? possibili riferimenti errati a \string\sref{#1} !!!}\fi
    \expandafter\xdef\csname @s@#1\endcsname{\la@sezionecorrente}
    \immediate\write16{\la@sezionecorrente. #2}
    \if@simboli
      \immediate\write\file@simboli{ }\immediate\write\file@simboli{ }
      \immediate\write\file@simboli{  Sezione
                                  \la@sezionecorrente :   sref.   #1}
      \immediate\write\file@simboli{ } \fi
    \if@riferimenti
      \immediate\write\file@ausiliario{\string\expandafter\string\edef
      \string\csname\b@lank @s@#1\string\endcsname{\la@sezionecorrente}}\fi
    \goodbreak\vskip 48pt plus 60pt
    \noindent{\bf\the\sect@num.\quad #2}\par\nobreak\vskip 15pt
    \nobreak\noindent}

\def\semiautosez#1#2\par{
    \gdef\la@sezionecorrente{#1}\if@ndoppia\global\eqnum=0\fi
    \if@simboli
      \immediate\write\file@simboli{ }\immediate\write\file@simboli{ }
      \immediate\write\file@simboli{  Sezione ** : sref.
          \expandafter\spo@glia\meaning\la@sezionecorrente}
      \immediate\write\file@simboli{ }\fi
    \s@ection#2\par}


\def\eqref#1{\se@indefinito{@eq@#1}
    \immediate\write16{ ??? \string\eqref{#1} non definita !!!}
    \expandafter\xdef\csname @eq@#1\endcsname{??}
    \fi\csname @eq@#1\endcsname}

\def\eqlabel#1{\global\advance\eqnum by 1
    \if@ndoppia\xdef\il@numero{\la@sezionecorrente.\the\eqnum}
       \else\xdef\il@numero{\the\eqnum}\fi
    \def\usa@getta{1}\se@indefinito{@eq@#1}\def\usa@getta{2}\fi
    \expandafter\ifx\csname @eq@#1\endcsname\il@numero\def\usa@getta{2}\fi
    \ifodd\usa@getta\immediate\write16
       { ??? possibili riferimenti errati a \string\eqref{#1} !!!}\fi
    \expandafter\xdef\csname @eq@#1\endcsname{\il@numero}
    \if@ndoppia
       \def\usa@getta{\expandafter\spo@glia\meaning
       \la@sezionecorrente.\the\eqnum}
       \else\def\usa@getta{\the\eqnum}\fi
    \if@simboli
       \immediate\write\file@simboli{  Equazione
            \usa@getta :  eqref.   #1}\fi
    \if@riferimenti
       \immediate\write\file@ausiliario{\string\expandafter\string\edef
       \string\csname\b@lank @eq@#1\string\endcsname{\usa@getta}}\fi}

\def\autoeqno#1{\eqlabel{#1}\eqno(\csname @eq@#1\endcsname)}
\def\autoleqno#1{\eqlabel{#1}\leqno(\csname @eq@#1\endcsname)}
\def\eqrefp#1{(\eqref{#1})}


\newcount\cit@num\global\cit@num=0

\newwrite\file@bibliografia
\newif\if@bibliografia
\@bibliografiafalse

\def\lp@cite{[}
\def\rp@cite{]}
\def\trap@cite#1{\lp@cite #1\rp@cite}
\def\lp@bibl{[}
\def\rp@bibl{]}
\def\trap@bibl#1{\lp@bibl #1\rp@bibl}

\def\refe@renza#1{\if@bibliografia\immediate        
    \write\file@bibliografia{
    \string\item{\trap@bibl{\cref{#1}}}\string
    \bibl@ref{#1}\string\bibl@skip}\fi}

\def\ref@ridefinita#1{\if@bibliografia\immediate\write\file@bibliografia{
    \string\item{?? \trap@bibl{\cref{#1}}} ??? tentativo di ridefinire la
      citazione #1 !!! \string\bibl@skip}\fi}

\def\bibl@ref#1{\se@indefinito{@ref@#1}\immediate
    \write16{ ??? biblitem #1 indefinito !!!}\expandafter\xdef
    \csname @ref@#1\endcsname{ ??}\fi\csname @ref@#1\endcsname}

\def\c@label#1{\global\advance\cit@num by 1\xdef            
   \la@citazione{\the\cit@num}\expandafter
   \xdef\csname @c@#1\endcsname{\la@citazione}}

\def\bibl@skip{\vskip 0truept}


\def\stileincite#1#2{\global\def\lp@cite{#1}\global
    \def\rp@cite{#2}}
\def\stileinbibl#1#2{\global\def\lp@bibl{#1}\global
    \def\rp@bibl{#2}}

\def\citpreset#1{\global\cit@num=#1
    \immediate\write16{ !!! cit-preset = #1 }    }

\def\autobibliografia{\global\@bibliografiatrue\immediate
    \write16{ !!! Genera il file \jobname.BIB}\immediate
    \openout\file@bibliografia=\jobname.bib}

\def\cref#1{\se@indefinito                  
   {@c@#1}\c@label{#1}\refe@renza{#1}\fi\csname @c@#1\endcsname}

\def\cite#1{\trap@cite{[\cref{#1}]}}                  
\def\ccite#1#2{\trap@cite{[\cref{#1},\cref{#2}]}}     
\def\ncite#1#2{\trap@cite{[\cref{#1}--\cref{#2}]}}    
\def\upcite#1{$^{\,\trap@cite{[\cref{#1}]}}$}               
\def\upccite#1#2{$^{\,\trap@cite{[\cref{#1},\cref{#2}]}}$}  
\def\upncite#1#2{$^{\,\trap@cite{[\cref{#1}-\cref{#2}]}}$}  
\def\Rcite#1{Ref.~[\cref{#1}]}

\def\clabel#1{\se@indefinito{@c@#1}\c@label           
    {#1}\refe@renza{#1}\else\c@label{#1}\ref@ridefinita{#1}\fi}

\def\biblskip#1{\def\bibl@skip{\vskip #1}}           

\def\insertbibliografia{\if@bibliografia             
    \immediate\write\file@bibliografia{ }
    \immediate\closeout\file@bibliografia
    \catcode`@=11\input\jobname.bib\catcode`@=12\fi}


\def\commento#1{\relax}
\def\biblitem#1#2\par{\expandafter\xdef\csname @ref@#1\endcsname{#2}}


\catcode`@=12

%
%
\centerline{\bm Third-order density-functional perturbation theory:}
\centerline{\bm a practical implementation with applications}
\centerline{\bm to anharmonic couplings in Si}

\bigskip
\centerline{\bf Alberto Debernardi and Stefano Baroni}
\smallskip
\centerline{\it Scuola Internazionale Superiore di Stud\^\i\ Avanzati (SISSA)}
\centerline{\it Via Beirut 2/4, I-34014 Trieste, Italy}
\vskip -10pt $$\vbox{\hsize=6truein \noindent \tiny\baselineskip=10pt
We present a formulation of
third-order density-functional perturbation theory which is manifestly
invariant with respect to unitary transfomations within the
occupied-states manifold and is particularly suitable for a practical
implementation of the so called `2n+1' theorem. Our implementation is
demonstrated with the calculation of the third-order anharmonic
coupling coefficients for some high-simmetry phonons in Silicon. } $$

\smallskip


Preprint babbage: cond-mat/9406063
\smallskip

\autobibliografia\stileincite{}{}

Density Functional Perturbation Theory (DFPT) is a powerful tool to
determine low-order derivatives of the ground-state electronic energy
of materials with respect to some external parameters. The use of DFPT
is twofold. On the one hand, it allows to calculate response
functions---which are directly accessible to experi\-ments---or other
measurable properties which can be related to response functions---such
as e.g. phonon frequencies in the adiabatic approximation. On the other
hand, it can be used to calculate properties of specific, complex,
materials which can viewed as small perturbations with respect to
other, simpler, systems. Since DFPT was demonstrated to be a
computationally viable technique \cite{BGT}, many applications have
appeared belonging to either categories. The first group includes the
calculation of elastic constants \cite{BGT2}, dielectric and
piezoelectric constants \cite{piezo}, and various lattice-dynamical
properties \ccite{GdGPB}{others}. Other applications belonging to the
second group are based on the so called {\it computational
alchemy} approach to semiconductor alloys \cite{alloys} and
superlattices \cite{offsets} in which the disordered semiconductor is
viewed as a small perturbation with respect to a reference, periodic
system (the {\it virtual} crystal).

All the applications of DFPT appeared so far are limited to second
order in the energy. It is a well known result of elementary
quantum-mechanics that the knowledge of the wavefunction response of a
system up to $n$-th order in the strength of an external perturbation
is sufficient to determine the energy derivative up to order $2n+1$
\cite{2n+1}. The validity of this `$2n+1$ theorem' within
self-consistent field (SCF) theories has been known since several years
in the quantum-chemistry community \cite{QC}, and recently it has been
generalized to density-functional theory (DFT) by Gonze and Vigneron
\cite{Gonze}. A first important conclusion we can draw from this
`theorem' is that the knowledge of the {\it linear} response of a
system to an external perturbation allows to determine the {\it third}
derivatives of the energy with respect to the strength of the
perturbation and it gives therefore a practical way to link linear and
quadratic generalized susceptibilities. The interest in doing so is
evident: one can in principle obtain higher-order susceptibilities or
gain in the accuracy achieved by perturbation theory essentially {\it
for free}. In the following, we will concentrate on the formulation by
Gonze and Vigneron and will restate it in a form which is free from
some of its original drawbacks, and is well suited for practical
implementations. As an example, we calculate the third-order anharmonic
coupling coefficients in Silicon at some high-symmetry points of the
Brillouin zone (BZ), and compare them with results obtained by the
frozen-phonon method.

Let us suppose that the {\it external} potential acting on the
electrons of a given system depends {\it linearly} on some external
parameter, $\lambda$: $$v_{ext}(\rr) \equiv v^0_{ext}(\rr) + \lambda
v^1_{ext}(\rr). \autoeqno{vext} $$ The case where the dependence of
$v_{ext}^\lambda$ upon $\lambda$ is non linear requires a
straightforward generalization of the results obtained in the linear
case and will be considered later. According to Gonze and Vigneron
\cite{Gonze}, the third-order derivative of the DFT ground-state energy
with respect to $\lambda$ reads: $$\displaylines{ \quad
\left . {\partial^3 E^\lambda
\over\partial \lambda^3} \right |_{\lambda=0} = 6 \sum_v \langle
\psi^1_v | H^1_{SCF} - \epsilon_v^1 | \psi^1_v \rangle +
\int K^3(\rr,\rr',\rr'') n^1(\rr) n^1(\rr') n^1(\rr'') d\rr
d\rr' d\rr'', \eqlabel{gonze_3} \quad\eqrefp{gonze_3} }$$ where the sum
runs over occupied ({\it valence}) Kohn-Sham (KS) orbitals, $\psi^1$
and $\epsilon^1$ indicate the first derivatives of the KS orbitals and
energy levels respectively, $n^1$ and $H^1_{SCF}$ indicate the
corresponding linear corrections to the electron ground-state density
and KS one-electron hamiltonian, and $K^3$ is finally the third-order
functional derivative of the exchange-correlation energy with respect
to the electron density: $K^3(\rr,\rr',\rr'') = \left . {\delta^3
E_{XC}[n] \over \delta n(\rr) \delta n(\rr') \delta n(\rr'')} \right
|_{n=n^0} $. Eq. \eqrefp{gonze_3} clearly shows that the calculation of
the third-order correction to the energy requires only the knowledge of
such ingredients as $\psi^1$ and $\epsilon^1$ which are directly
accessible to first-order perturbation theory.

All the results of DFT must be invariant with respect to unitary
transformations of the orbitals which do not mix the manifolds of
occupied and empty ({\it conduction}) states. As it stands, Eq.
\eqrefp{gonze_3} does not manifestly display this invariance.
Furthermore, its implementation would require the knowledge of the
components of the perturbed wavefunction, $\psi^1_v$, along all the
valence wavefunctions different from $\psi^0_v$ itself: $\langle
\psi^0_{v'} | \psi^1_{v} \rangle_{v' \ne v} $. Once again, this is
innatural because in DFT the variation of any physical property must
only depend on the variation of the one-electron density matrix which
is not affected by components of the perturbed valence orbitals along
the unperturbed valence manifold. This situation is particularly
unpleasant when, due to the degeneracy or quasi-degeneracy of some
unperturbed valence states, the actual implementation of Eq.
\eqrefp{gonze_3} would require the use of degenerate-state perturbation
theory. We will now show that Eq. \eqrefp{gonze_3} can be recast in a
form which requires only the knowledge of the conduction-manifold
projection of the $\psi^1_v$'s, which is manifestly invariant with
respect to unitary transformations within the valence manifold, and
which can be straightforwardly and efficiently implemented using
standard non-degenerate first-order perturbation theory.

The second term on the right-hand side (rhs) of Eq. \eqrefp{gonze_3}
already displays the desired unitary invariance, and we concentrate on
the first term. Our final result is: $$\sum_v
\langle \psi^1_v | H^1_{SCF} - \epsilon_v^1 | \psi^1_v \rangle = \sum_v
\langle \psi^1_v | P_c H_{SCF}^1 P_c | \psi^1_v \rangle -
\sum_{vv'} \langle \psi_v^1 | P_c | \psi_{v'}^1 \rangle \langle
\psi_{v'}^0 | H^1_{SCF} | \psi_v^0 \rangle, \autoeqno{alberto_3}
$$ where $P_c \equiv \sum_c
|\psi_c^0\rangle\langle \psi_c^0 | $ is the projector over the
unperturbed conduction-state manifold (from now on, `$c$' will indicate
an index running over conduction states, while `$v$' indicates sums
over valence states). Before demonstrating Eq. \eqrefp{alberto_3} we
notice that it is manifestly invariant with respect to unitary
transformations within the valence manifold. In fact, it is the sum of
the trace of a matrix defined over that manifold (first term on the
rhs) and of the product of two such matrices (second term). The desired
invariance derives from the invariance property of the trace. The
demonstration of Eq. \eqrefp{alberto_3} is tedious, but
straightforward. Let us start from the definition of the first-order
correction to the $v$-th unperturbed valence state, and consider its
projections over the valence- and conduction-state manifolds: $$
\eqlabel{psi1} \displaylines{ \hfill | \psi^1_v \rangle = P_c |
\psi^1_v \rangle + P_v | \psi^1_v \rangle, \hfill
\llap{(\eqref{psi1}a)} \cr \hfill P_c | \psi^1_v \rangle = \sum_{c}
|\psi^0_c \rangle { \langle \psi^0_c | H_{SCF}^1 | \psi^0_v \rangle
 \over \epsilon^0_v - \epsilon^0_c }, \hfill \llap{(\eqref{psi1}b)} \cr
\hfill P_v | \psi^1_v \rangle = \sum_{v'\ne v} |\psi^0_{v'} \rangle {
\langle \psi^0_{v'} | H_{SCF}^1 | \psi^0_v \rangle
 \over \epsilon^0_v - \epsilon^0_{v'} }, \hfill \llap{(\eqref{psi1}c)}
} $$ where $P_v \equiv 1-P_c$ is the projector over the valence
manifold.
Substituting Eq. (\eqref{psi1}a) into the left-hand size of Eq.
\eqrefp{alberto_3},
one obtains the sum of four terms, which will be denoted
by $cc$, $cv$, $vc$, and $vv$, according to the couple of projectors
appearing inside the matrix elements. Inserting Eq. (\eqref{psi1}c)
into the expression of the $vv$ term and separating out terms with
$v'=v''$ from those with $v'\ne v''$, one obtains:
$$\eqlabel{vv}\displaylines{\quad \sum_v \langle \psi^1_v |P_v(
H^1_{SCF} - \epsilon_v^1) P_v | \psi^1_v \rangle =
\sum_{v\ne v'} { \langle \psi_v^0 | H^1_{SCF} | \psi_{v'}^0 \rangle
(\epsilon^1_{v'} - \epsilon^1_{v} ) \langle \psi^0_{v'} | H^1_{SCF} |
\psi^0_v \rangle \over (\epsilon^0_v - \epsilon^0_{v'})^2 } +
\hfill \cr \hfill +
\sum_{v\ne v'\ne v''} { \langle \psi_v^0 | H^1_{SCF} | \psi_{v'}^0
\rangle \langle \psi_{v'}^0 | H^1_{SCF} | \psi_{v''}^0 \rangle \langle
\psi_{v''}^0 | H^1_{SCF} | \psi_v^0 \rangle \over (\epsilon^0_v -
\epsilon^0_{v'}) (\epsilon^0_v - \epsilon^0_{v''})}.
\quad
\eqrefp{vv} }$$ Both the terms on the rhs of Eq. \eqrefp{vv} vanish
because the parities of the numerators and those of the denominators
with respect to the exchanges $v\rightleftharpoons v'$
$v\rightleftharpoons v''$ are different. Let us come now to the $cv$
and $vc$ terms. Using Eqs. (\eqref{psi1}b) and (\eqref{psi1}c) and a
few algebraic manipulations,
one obtains: {\overfullrule=0pt $$\eqlabel{cv}\displaylines{
\sum_v \Bigl ( \langle \psi^1_v |P_c( H^1_{SCF} - \epsilon_v^1) P_v |
\psi^1_v \rangle+ \langle
\psi^1_v |P_v( H^1_{SCF} - \epsilon_v^1) P_c | \psi^1_v \rangle \Bigr )
= \cr \hfill \sum_{c,v\ne v'}{ \langle
\psi_v^0 | H^1_{SCF} | \psi^0_{c} \rangle \langle \psi_{c}^0 |
H^1_{SCF} | \psi^0_{v'} \rangle \langle \psi_{v'}^0 | H^1_{SCF} |
\psi^0_v \rangle \over (\epsilon^0_v - \epsilon^0_{v'}) }
\left ( {1\over
\epsilon_v^0-\epsilon_c^0} - {1\over \epsilon_{v'}^0-\epsilon_c^0}
\right ) = \hfill \cr \hfill - \sum_{c,v\ne v'}{ \langle
\psi_v^0 | H^1_{SCF} | \psi^0_{c} \rangle \langle \psi_{c}^0 |
H^1_{SCF} | \psi^0_{v'} \rangle \langle \psi_{v'}^0 | H^1_{SCF} |
\psi^0_v \rangle \over (\epsilon^0_v - \epsilon^0_{c}) (\epsilon^0_{v'}
- \epsilon^0_{c})}.
\hfill\llap{\eqrefp{cv}}} $$} The $cc$ term reads:
$$\eqlabel{cc}\displaylines{\quad \sum_v \langle \psi^1_v |P_c(
H^1_{SCF} - \epsilon_v^1) P_c | \psi^1_v \rangle = \hfill \cr \hfill
\sum_v \langle \psi^1_v |P_c H^1_{SCF} P_c | \psi^1_v \rangle
- \sum_{c,v}{ \langle \psi_v^0 | H^1_{SCF} | \psi^0_{c}
\rangle \langle \psi_{c}^0 | H^1_{SCF} | \psi^0_{v} \rangle \langle
\psi_{v}^0 | H^1_{SCF} | \psi^0_v \rangle \over (\epsilon^0_v -
\epsilon^0_{c}) (\epsilon^0_{v} - \epsilon^0_{c})}. \quad
\eqrefp{cc} } $$ The first term on the
rhs of Eq. \eqrefp{cc} coincides with the first term on the rhs of Eq.
\eqrefp{alberto_3}. The second term has the same form as the rhs of
Eq. \eqrefp{cv}, just providing the $v=v'$ terms which were missing
therein. By combining these terms, we finally obtain:
$$\eqlabel{penultima}\displaylines{\quad \sum_v \langle \psi^1_v |
H^1_{SCF} - \epsilon_v^1 | \psi^1_v \rangle = \sum_v \langle \psi^1_v |
P_c H_{SCF}^1 P_c | \psi^1_v \rangle - \hfill \cr \hfill - \sum_{c,v}{
\langle \psi_v^0 | H^1_{SCF} | \psi^0_{c} \rangle \langle \psi_{c}^0 |
H^1_{SCF} | \psi^0_{v'} \rangle \langle \psi_{v'}^0 | H^1_{SCF} |
\psi^0_v \rangle \over (\epsilon^0_v - \epsilon^0_{c}) (\epsilon^0_{v'}
- \epsilon^0_{c})}. \quad
 \eqrefp{penultima}} $$ By using Eq. (\eqref{psi1}b) and the condition
that different conduction state are orthogonal to each other, we
finally arrive at Eq. \eqrefp{alberto_3}.

Eq. \eqrefp{alberto_3} can be easily generalized to the case where the
perturbation depends nonlinearly on more than one parameter (as it is
actually the case, e.g., in lattice dynamics if $\lambda$ is identified
with a nuclear displacement). Suppose there are three different such
parameters, $\{\lambda_1,\lambda_2,\lambda_3\}$. Following the notation
and the line of reasoning of \Rcite{Gonze}, we easily arrive at the
final result: $$ {\partial^3 E \over
\partial\lambda_1 \partial\lambda_2 \partial\lambda_3} = \widetilde
E^{\lambda_1\lambda_2\lambda_3} + \widetilde
E^{\lambda_2\lambda_1\lambda_3} + \widetilde
E^{\lambda_1\lambda_3\lambda_2} + \widetilde
E^{\lambda_3\lambda_1\lambda_2} + \widetilde
E^{\lambda_3\lambda_2\lambda_1} + \widetilde
E^{\lambda_2\lambda_3\lambda_1} ,\autoeqno{d3e}$$ where
$$\displaylines{\eqlabel{Elll} \quad \widetilde
E^{\lambda_1\lambda_2\lambda_3} = \hfill \cr \noalign{\vskip 3pt}
\hfill \sum_v \Bigl (
\langle\psi_v^{\lambda_1}|P_cH_{SCF}^{\lambda_2}P_c| \psi_v^{\lambda_3}
\rangle
+ \langle\psi_v^{\lambda_1}|P_c
v^{\lambda_2\lambda_3} | \psi^0_v \rangle + \langle
\psi^0_v|v^{\lambda_1\lambda_2} P_c | \psi_v^{\lambda_3} \rangle
+\langle\psi_v^0 |
v^{\lambda_1\lambda_2\lambda_3} | \psi_v^0 \rangle \Bigr )
\hfill\cr\hfill
- \sum_{vv'}
\langle \psi_v^{\lambda_1} | P_c | \psi_{v'}^{\lambda_2} \rangle
\langle \psi^0_{v'} | H_{SCF}^{\lambda_3} | \psi_v^0 \rangle
+ {1 \over 6} \int K^3(\rr,\rr',\rr'') n^{\lambda_1}(\rr)
n^{\lambda_2}(\rr') n^{\lambda_3}(\rr'') d\rr d\rr' d\rr'',\hfill
\eqrefp{Elll} } $$ where the $\lambda_i$ superscript indicates the
derivative with respect to $\lambda_i$. One sees that when the external
potential depends lineary on just one parameter, $\lambda$, Eq.
\eqrefp{gonze_3} is recovered. In the general case where the positions
of the nuclei also depend on the $\lambda$'s, one must of course add to
Eq. \eqrefp{d3e} the derivative of the ionic contribution to the energy
which is usually expressed as an Ewald sum. All the ingredients
necessary to implement Eq. \eqrefp{Elll} are naturally provided by any
computer code aimed at standard second-order DFPT, such as the one we
routinely use for lattice-dynamical calculations. In the following, we
present some tests of the above formulation which we have made on the
anharmonic
coupling between lattice distortions of Silicon at selected
high-simmetry points of the BZ.

The equilibrium and lattice-dynamical properties of Silicon have been
calculated
within the local-density approximation, using the plane-wave
pseudopotential method. We have used the same pseudopotential as in
\Rcite{GdGPB}, plane waves up to a kinetic-energy cutoff of 14 Ry,
and the (444) Monkhorst-Pack mesh for BZ integrations \cite{MP}.
Calculations have been done
at the $\Gamma$ and $X$ points of the BZ
both within DFPT and, for comparison, by the frozen-phonon method. In the
latter case, the unit cell has a lower (rotational and/or translational)
symmetry, and the set of $\bf k$-point used for sampling the BZ has been
modified accordingly. We stress that, as it is the case for the harmonic
dynamical matrix \cite{GdGPB}, the calculation of anharmonic
coefficients at arbitrary points of the BZ within DFPT does {\it not}
require the use of any supercells, but it only uses wavefunctions and
band energies calculated for the unperturbed system. There are four
independent parameters describing the harmonic properties of the crystal
within the set of distortions corresponding to $\Gamma$ and $X$ phonons
(the $\Gamma_{LTO}$, $X_{LAO}$, $X_{TA}$, and $X_{TO}$ frequencies), whereas
there are six anharmonic constants: one describing the coupling between
three $\Gamma$-like phonons, and five describing the coupling between one
$\Gamma$- and two $X$-like phonons. We refer to \Rcite{VdB} for a
full group-theoretical analysis of the independent coupling coefficients
and for an explanation of the notations we borrow from it. In Table I
we compare the third-order coupling coefficients calculated in the
present work with DFPT and the frozen-phonon method. The values obtained
with the latter method have been obtained using a procedure analogous to
the one used in \Rcite{VdB}.
As one can see, DFPT give results which are
in perfect agreement with those obtained by the frozen-phonon
method. Actually, they are in principle more accurate because DFPT
directly provides the energy derivatives without the need of any
numerical differentiations.
\midinsert
{\tolerance 100000
\baselineskip=10pt\noindent
\def \tabrule     {\noalign{\vskip 5truept \hrule\vskip 5truept} }
\def \tabrul2     {\noalign{\vskip 5truept \hrule \vskip 2truept \hrule
                   \vskip 5truept} }
$$
\setbox0=\vbox{\halign {
 # \hfil &
\hfil \quad # \quad \hfil &
\hfil \quad # \quad \hfil &
\hfil \quad # \quad \hfil &
\hfil \quad # \quad \hfil &
\hfil \quad # \quad \hfil &
\hfil \quad # \quad \hfil \cr
\tabrul2
 & $B_{xyz} $ & $I_{z\overline{aa}} $ & $I_{z\overline{bb}}
$ & $I_{z\overline{cc}} $ &
$I_{\overline{xac}} $ & $I_{\overline{ybc}}
$ \cr
\tabrule
DFPT & 295.06 & 232.41 & -35.27 & 55.92 & 447.64 & -64.74 \cr
FP   & 295.27 & 232.11 & -35.23 & 55.44 & 447.19 & -64.84 \cr
\tabrul2
}}
\vbox{\hsize = \wd0
\vbox{\noindent TABLE I. Comparison of the third-order anharmonic coupling
constants between phonons at the $\Gamma$ and $X$ points of the
Brillouin zone in Silicon, as obtained by density-functional
perturbation theory (DFPT) and the frozen-phonon (FP) method. The
notations are the same as in
Ref. [12]. Units are $\rm
eV/\AA^3$.}
\box0}
$$}
\endinsert
We conclude that DFPT provides an accurate and computationally
convenient tool for calculating the anharmonic coupling of phonons at
arbitrary points of the BZ, with a numerical effort which essentially
does not depend on the position in the BZ. This opens the way to a
systematic investigation of such effects in real materials. A
calculation of the anharmonic-decay phonon lifetimes in semiconductors
along the lines presented in this paper is presently under way
\cite{DBBM}.

\medskip We are grateful to S. de Gironcoli, E. Molinari, and R. Resta
for frequent and useful discussions.

\def\prb#1#2#3{Phys. Rev. B {\bf #1}, #2 (#3)}
\def\prl#1#2#3{Phys. Rev. Lett. {\bf #1}, #2 (#3)}

\biblitem{MP} H.~J.~Monkhorst and J.~D.~Pack, \prb{13}{5188}{1976}.

\biblitem{DBBM} A. Debernardi, S. Baroni, and E. Molinari, unpublished.

\biblitem{VdB} D.~Vanderbilt S.~H.~Taole and S.~Narasimhan,
Phys. Rev. B {\bf 40}, 5657 (1989).

\biblitem{2n+1} See for instance P.~M.~Morese and H.~Feshbach, {\it
Methods of Theoretical Physics} (Mc Graw-Hill, New York, 1953), Vol. II,
P. 1120.

\biblitem{QC} T.~S.~Nee, R.~G.~Parr and R.~J.~Bartlett, J.~Chem. Phys.
 {\bf 64}, 2216 (1976)

\biblitem{piezo} S. de Gironcoli, S. Baroni, and R. Resta,
\prl{62}{2843}{1989}; S. de Gironcoli, S. Baroni, and R. Resta,
Ferroelectrics, {\bf 111}, 19 (1990); A. Dal Corso, S. Baroni, and R.
Resta, \prb{47}{3588}{1993}; A. Dal Corso, S. Baroni, and R.  Resta,
\prb{49}{5323}{1994}.

\biblitem{Gonze} X. Gonze and J.-P. Vigneron, \prb{39}{13120}{1989}.

\biblitem{BGT} S. Baroni, P. Giannozzi, and A. Testa,
\prl{58}{1861}{1987}.

\biblitem{GdGPB} P.~Giannozzi, S.~de~Gironcoli, P.~Pavone, and S.~Baroni,
\prb{43}{7231}{1991}.

\biblitem{BGT2} S. Baroni, P. Giannozzi, and A. Testa,
\prl{59}{2662}{1987}.

\biblitem{others} F.~Ancilotto, A.~Selloni, W.~Andreoni, S.~Baroni,
R.~Car, and M.~Parrinello, \prb{43}{8930}{1991}; X. Gonze, D.~C.~Allan,
and M.~P.~Teter, \prl{68}{3603}{1992}; M. Buongiorno Nardelli,
S.~Baroni, and P.~Giannozzi, \prl{69}{1069}{1992}; A. Dal Corso, S.
Baroni, R. Resta, and S. de Gironcoli, \prb{47}{3588}{1993}; P. Pavone,
K. Karch, O. Sch\"utt, D. Strauch, P. Giannozzi, and S. Baroni,
\prb{48}{3156}{1993}; J. Fritsch, P. Pavone, and U. Schr\"oder,
\prl{71}{4194}{1993}; C.~Lee and X.~Gonze, \prl{72}{1686}{1994};
C.-Z.~Wang, R.~Yu, and H.~Krakauer, \prl{72}{368}{1994};
S~.Y.~Savranov, D.~Y.~Svranov, and O.~K.~Andersen, \prl{72}{372}{1994};
P. Giannozzi and S. Baroni, J. Chem. Phys. {\bf 100}, 8537 (1994).

\biblitem{alloys} S. de Gironcoli, P. Giannozzi, and S. Baroni,
\prl{66}{2196}{1991};
N. Marzari, S. de Gironcoli, and S. Baroni, \prl{72}{4011}{1994}.

\biblitem{offsets}
S. Baroni, M. Peressi, R. Resta, and A. Baldereschi, Proc. 21$^{st}$
Intl. Conf. on {\it The Physics of Semiconductors}, edited by P. Jiang
and Hou-Zhi Zheng (World-Scientific, Singapore, 1993), p. 863;
M. Pe\-res\-si and S. Baroni, \prb{49}{7490}{1994}.

\bigskip\centerline{REFERENCES}\medskip

\insertbibliografia

\vfill\eject\bye

\vfill\eject

\bye